\documentclass[twocolumn,prl,letterpaper,showpacs]{revtex4}

\usepackage{amsmath}
\usepackage{graphicx}
\usepackage{amsthm}
\usepackage{amsfonts}
\usepackage{natbib}

\newtheorem*{Theorem}{Theorem}
\newtheorem{Lemma}{Lemma}
\newtheorem*{Corollary}{Corollary}
\newcommand{\Ket}[1]{\ensuremath{\left | #1 \right \rangle}}
\newcommand{\Bra}[1]{\ensuremath{\left \langle #1 \right |}}

\begin{document}

\title{Pre- and Post-selection paradoxes and contextuality in quantum mechanics}
\author{M. S. Leifer}
\email{mleifer@perimeterinstitute.ca}
\author{R. W. Spekkens}
\email{rspekkens@perimeterinstitute.ca} \affiliation{Perimeter
Institute for Theoretical Physics, 31 Caroline Street North,
Waterloo, Ontario, Canada, N2L 2Y5}

\begin{abstract}
Many seemingly paradoxical effects are known in the predictions
for outcomes of intermediate measurements made on pre- and
post-selected quantum systems. Despite appearances, these effects
do not demonstrate the impossibility of a noncontextual hidden
variable theory, since an explanation in terms of
measurement-disturbance is possible. Nonetheless, we show that for
every paradoxical effect wherein all the pre- and post- selected
probabilities are 0 or 1 and the pre- and post-selected states are
nonorthogonal, there is an associated proof of contextuality.
This proof is obtained by considering all the measurements
involved in the paradoxical effect -- the pre-selection, the
post-selection, and the alternative possible intermediate
measurements -- as alternative possible measurements at a single
time.
\end{abstract}

\maketitle

The study of quantum systems that are both pre- and post-selected
was initiated by Aharonov, Bergmann and Lebowitz (ABL) in 1964
\cite{ABLRule}, and has led to the discovery of many
counter-intuitive results, which we refer to as \emph{pre- and
post-selection (PPS) effects} \cite{AhaVaidReview}, some of which
have recently been confirmed experimentally
\cite{LundeenReschSteinberg}.

These results have led to a long debate about the interpretation
of the ABL probability rule \cite{Kastner7}. An undercurrent in
this debate has been the connection between PPS effects and
contextuality.  For instance, Bub and Brown \cite{BubBrown}
understood the paper of Albert, Aharonov and D'Amato
\cite{Proto3Box} -- which concerned a PPS effect known since as
the ``3-box paradox'' -- as a claim to a novel proof of
contextuality, that is, as a version of the Bell-Kochen-Specker
theorem \cite{BellKS, KS}, and convincingly disputed this claim.
Although the language of Ref.\ \cite{Proto3Box} does suggest such
a reading, in Ref. \cite{AADcomment} the authors clarified their
position, stating that it was not their intention to conclude
anything about hidden variable theories. Nonetheless, discussions
of PPS paradoxes, that is, PPS effects of the 3-box paradox
variety, continue to make use of a language that suggests
implications for ontology \cite{VaidTS3} and the claim that
certain PPS paradoxes \emph{are} proofs of contextuality can be
found in the literature \cite{Marchildon}. Although we agree with
Bub and Brown that this claim is mistaken, we show that there is
nonetheless a close connection between the two phenomena.

This connection is expected to have interesting applications in
quantum foundational studies. For instance, it has been suggested
by some that Bell's theorem \cite{Bell} might be understood within
a realist and Lorentz-invariant framework if one admits the
possibility of a hidden variable theory that allows for
backward-in-time causation \cite{Price}.  A simple model has even
been suggested by Kent \cite{Kentrevisited}. The latter is closely
connected to the fact that Bell correlations can be simulated
using postselection, as shown in Bub and Brown \cite{BubBrown}.
This simulation by postselection is also the root of the
detection-efficiency loophole in experimental tests of Bell's
theorem \cite{Fine,Pearle}. Further investigations into the
connection between proofs of nonlocality and PPS paradoxes would
shed new light on these avenues of research.  As nonlocality is a
kind of contextuality (assuming separability
\cite{SpekkensContPMT}), the ubiquitous connection between
contextuality and PPS paradoxes established in the present work is
an important contribution to this project. Moreover, the fact that
the phenomenon of contextuality \emph{itself} might be understood
by abandoning the traditional notion of causality, and the fact
that its simulation by postselection will likely constitute a
loophole for experimental tests of contextuality, makes this
connection interesting in its own right.

 Mermin \cite{MerminMagic} has already shown
one connection between PPS effects and contextuality. His
investigation concerned what is known as the ``mean king's
problem'' \cite{MeanKing1} which is a PPS effect that is
qualitatively different from the paradoxical variety of PPS effect
that we shall be considering.  Moreover, Mermin demonstrated how
one can obtain a type of mean king's problem that is unsolvable
starting from the measurements used in a proof of contextuality,
whereas we demonstrate how one can obtain proofs of contextuality
starting from the measurements used in a PPS paradox.

To be specific, we demonstrate the following: for every PPS
paradox wherein all the PPS probabilities are 0 or 1 and the pre-
and post-selection states are non-orthogonal, there is an
associated proof of contextuality. The key to the proof is that
measurements that are treated as temporal successors in the PPS
paradox are treated as counterfactual alternatives in the proof of
contextuality. This result suggests the existence of a subtle
conceptual connection between the two phenomena that has yet to be
fully understood.  Thus, the present work contributes to the
project of reducing the number of logically distinct quantum
mysteries by revealing the connections between them.

We begin with a curious prediction of the ABL rule known as the
\emph{3-box paradox}. Suppose we have a particle that can be in
one of three boxes. We denote the state where the particle is in
box $j$ by $\left\vert j\right\rangle $. The particle is
pre-selected in the state $\left\vert \phi \right\rangle =
\left\vert 1\right\rangle +\left\vert 2\right\rangle +\left\vert
3\right\rangle $ and post-selected in the state $ \left\vert \psi
\right\rangle = \left\vert 1\right\rangle +\left\vert
2\right\rangle -\left\vert 3\right\rangle $ (states will be left
unnormalized). At an intermediate time, one of two possible
measurements is performed. The first possibility corresponds to
the projector valued measure (PVM)~\footnote{ A PVM is a set of
projectors that sum to the identity operator}
$\mathcal{E}_{1}=\{P_{1},P_{1}^{\perp }\}, $ where
\begin{equation}
P_{1}=\left\vert 1\right\rangle \left\langle 1\right\vert \qquad
P_{1}^{\perp }=\left\vert 2\right\rangle \left\langle 2\right\vert
+\left\vert 3\right\rangle \left\langle 3\right\vert .
\end{equation}%
The second possibility corresponds to the PVM
$\mathcal{E}_{2}=\{P_{2},P_{2}^{\perp }\},$ where
\begin{equation}
P_{2}=\left\vert 2\right\rangle \left\langle 2\right\vert \qquad
P_{2}^{\perp }=\left\vert 1\right\rangle \left\langle 1\right\vert
+\left\vert 3\right\rangle \left\langle 3\right\vert
\end{equation}
Now note that $P_1^{\perp}$ can also be decomposed into a sum of
the projectors onto the vectors $|2\rangle+|3\rangle$ and
$|2\rangle-|3\rangle$. However, the first of these is orthogonal
to the post-selected state, while the second is orthogonal to the
pre-selected state, so that the probability of the outcome
$P_1^{\perp}$ occurring, given that the pre- and post-selection
were successful, must be 0.  Consequently, the measurement of
$\mathcal{E}_{1}$ necessarily has the outcome $P_1$.  Similarly,
$P_2^{\perp}$ can be decomposed into a sum of the projectors onto
the vectors $|1\rangle+|3\rangle$ and $|1\rangle-|3\rangle$, which
are also orthogonal to the post- and pre-selected states
respectively. Consequently, the measurement of $\mathcal{E}_{2}$
necessarily has the outcome $P_2$. Thus, if one measures to see
whether or not the particle was in box $1$, one finds that it was
in box $1$ with certainty, and if one measures to see whether or
not it was in box $2$, one finds that it was in box $2$ with
certainty!

This is reminiscent of the sort of conclusion that one obtains in
proofs of the impossibility of a noncontextual hidden variable
theory.  Indeed, a proof presented by Clifton \cite{CliftonProof}
makes use of the same mathematical structure, as we presently
demonstrate.

\begin{figure}[h]
\includegraphics[width=80mm]{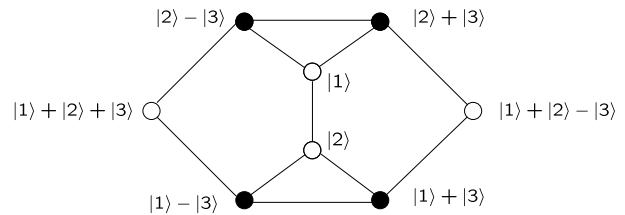}
\caption{A depiction of the vectors in Clifton's proof of
contextuality and their orthogonality relations. An assignment of
the value 1 or 0 to a vector is represented by colouring the
associated point white or black respectively.} \label{8rays}
\end{figure}

Consider the eight vectors mentioned explicitly in our discussion
of the 3-box paradox, but imagine that these describe alternative
possible measurements \emph{at a single time} (in contrast to what
occurs in the 3-box paradox). In a noncontextual hidden variable
theory, it is presumed that although not all of these tests can be
implemented simultaneously, their outcomes are determined by the
values of preexisting hidden variables and are independent of the
manner in which the test is made (the context). Thus each of these
vectors is assigned a value, 1 or 0, specifying whether the
associated test is passed or not. For any orthogonal pair, not
both can receive the value 1, and for any orthogonal triplet,
exactly one must receive the value 1. Representing the vectors by
points and orthogonality between vectors by a line between points,
these eight vectors above can be depicted as in Fig.\ \ref{8rays}.

Clifton's proof is an example of a \emph{probablistic} proof of
contextuality \cite{Cabello}, since it relies on assigning the
states $\Ket{\phi},\Ket{\psi}$ probability $1$ \emph{a priori}.
This is justified as follows: the state $\Ket{\phi}$ can be
prepared, and if it is, then a subsequent test for $\Ket{\phi}$
will be passed with certainty, and a subsequent test for
$\Ket{\psi}$ will be passed with nonzero probability (because
$\left\langle \psi \right\vert \left. \phi \right\rangle \neq 0$).
This implies that there must be some values of the hidden
variables that assign value 1 to both. Consider such a hidden
state. Since $|1\rangle-|3\rangle$ and $|2\rangle-|3\rangle$ are
orthogonal to $\left\vert \psi \right\rangle $, they must be
assigned value $0$ for this hidden state and since
$|1\rangle+|3\rangle$ and $|2\rangle+|3\rangle$ are orthogonal to
$\left\vert \phi \right\rangle $, they must also be assigned value
$0$. But given that $|1\rangle$, $|2\rangle+|3\rangle$, and
$|2\rangle-|3\rangle$ form an orthogonal triplet, it follows that
$|1\rangle$ must receive the value 1. Similarly, given that
$|2\rangle$, $|1\rangle+|3\rangle$, and $|1\rangle-|3\rangle$ form
an orthogonal triplet, it follows that $|2\rangle$ must receive
the value 1. However, $|1\rangle$ and $|2\rangle$ cannot both
receive the value 1, since they are orthogonal.  Thus, we have
derived a contradiction.

To our knowledge, the connection between Clifton's proof and the
3-box paradox has not previously been recognized.

We will demonstrate that this sort of connection is generic to PPS
paradoxes. We begin with a short review of the ABL rule, hidden
variable theories and contextuality.

We only consider quantum systems with a finite dimensional Hilbert
space and assume that no evolution occurs between measurements. We
restrict our attention to \emph{sharp} measurements, that is,
those associated with PVMs. We also restrict attention to
measurements for which the state updates according to $\rho
\rightarrow P_{j}\rho P_{j}/\text{Tr}(P_j \rho)$ upon obtaining
outcome $j$. This is known as the L{\"u}ders rule \cite{Luders}.
We call this set of assumptions the \emph{standard framework} for
PPS effects. It includes all of the PPS ``paradoxes'' discussed in
the literature to date.  The extent to which our results can be
generalized beyond this framework is a question for future
research.

To describe pre- and post-selected systems, we imagine a temporal
sequence of three sharp measurements. \ The initial, intermediate,
and final measurements occur at times $t_{\text{pre}}$, $t$, and
$t_{\text{post}}$ respectively, where $t_{\text{pre}} < t <
t_{\text{post}}$. The only relevant aspects of the initial and
final PVMs are the projectors associated with the outcomes
specified by the pre- and post-selection. Let these be denoted by
$\Pi _{\text{pre}}$ and $\Pi _{\text{post}}$ respectively, and let
the PVM associated with the intermediate measurement be denoted by
$\mathcal{E} = \{P_{j}\}$.

Assuming that nothing is known about the system prior to
$t_{\text{pre}},$ so that the initial density operator is $I/d,$
where $I$ is the identity operator, the measurement at
$t_{\text{pre}}$ prepares the density operator $ \rho
_{\text{pre}}=\Pi _{\text{pre}}/$Tr$(\Pi _{\text{pre}}).$ By
Bayes' theorem, we can deduce that the probability of obtaining
the outcome $k$ in the intermediate measurement is
\begin{equation*}
p(P_{k}|\Pi _{\text{pre}},\Pi
_{\text{post}},\mathcal{E})=\frac{\text{Tr}(\Pi
_{\text{post}}P_{k}\Pi _{\text{pre}}P_{k}))}{\sum_{j}\text{Tr}(\Pi _{\text{%
post}}P_{j}\Pi _{\text{pre}}P_{j})}.
\end{equation*}
This is a special case of the most general version of the ABL rule
\cite{AhaVaidReview}, and we therefore refer to such probabilities
as ``ABL probabilities''. In the case where $\Pi_{\text{pre}}$ and
$\Pi_{\text{post}}$ are rank-1 projectors onto states $\Ket{\phi}$
and $\Ket{\psi}$ respectively, this rule reduces to
\begin{equation}
p(P_k|\phi,\psi,\mathcal{E}) =
\frac{|\Bra{\psi}P_k\Ket{\phi}|^2}{\sum_j |\Bra{\psi}P_j
\Ket{\phi}|^2}
\end{equation}
which was implicitly used in our discussion of the 3-box paradox.
We now review hidden variable theories.

A hidden variable theory is an attempt to understand quantum
measurements as revealing features of pre-existing \emph{ontic
states}, by which we mean complete specifications of the state of
reality. A particularly natural class of such theories are those
that satisfy the following two assumptions \cite{SpekkensContPMT}:
\emph{measurement noncontextuality}, which is the assumption that
the manner in which the measurement is represented in the HVT
depends only on the PVM and not on any other details of the
measurement (the context); and \emph{outcome determinism for sharp
measurements}, which is the assumption that the outcome of a PVM
measurement is uniquely fixed by the ontic state. We abbreviate
these as MNHVTs. It follows that in an MNHVT, projectors are
associated with unique pre-existing properties that are simply
revealed by measurements \cite{SpekkensContPMT}.

Suppose we denote by $s$ the proposition that asserts that the
property associated with projector $P$ is possessed. \ In an MNHVT
the negation of $s,$ denoted $\lnot s,$ is associated with $I-P.$
\ Now consider a projector $Q$ that commutes with $P,$ and denote
the proposition associated with $Q$ by $t.$ In an MNHVT the
conjunction of $s$ and $t,$ denoted $s\wedge t,$ is associated
with $PQ$ and the disjunction of $s$ and $t,$ denoted $s\vee t,$ is
associated with $P+Q-PQ$ (the latter follows from the the fact
that $s\vee t=\lnot(\lnot s\wedge \lnot t)).$

Let $p(s)$ denote the probability that the proposition $s$ is
true. Classical probability theory dictates that
\begin{align}
0& \leq p(s)\leq 1 \\
p(\lnot s)& =1-p(s), \\
p(s\vee \lnot s)& =1,\;\text{ }p(s\wedge \lnot s)=0 \\
p(s\wedge t)& \leq p(s),\;p(s\wedge t)\leq p(t) \\
p(s\vee t)& =p(s)+p(t)-p(s\wedge t)
\end{align}
We therefore obtain the following constraints on an MNHVT.

\textbf{Algebraic conditions: }For projectors $P,Q$ such that
$[P,Q]=0$,
\begin{align}
0& \leq p(P)\leq 1  \label{ac0} \\
p(I-P)& =1-p(P),  \label{ac1} \\
p(I)& =1,\;\text{ }p(0)=0,  \label{ac2} \\
p(PQ)& \leq p(P),\;\text{ }p(PQ)\leq p(Q),  \label{ac3} \\
p(P+Q-PQ)& =p(P)+p(Q)-p(PQ).  \label{ac4}
\end{align}

By the assumption of outcome determinism, the probability assigned
to every projector for a particular ontic state is either 0 or 1.
The Bell-Kochen-Specker theorem shows that there are sets of
projectors to which no such assignment consistent with the
algebraic conditions is possible.

A connection to PPS paradoxes is suggested by the fact that there
exist sets of projectors for which an ABL\ probability assignment
violates the algebraic constraints, while every projector receives
probability 0 or 1. We call such a scenario a \emph{logical PPS
paradox}. The 3-box paradox is an example of this
~\footnote{Another example is the ``failure of the product rule''
discussed in \cite{VaidLorentz, CohenAntiABL}. By our results,
this is related to the contextuality proofs discussed in
\cite{MerminKS,PeresKS,KernaghanKS}.}.

Now, \emph{if} it were the case that in the HVT, the pre- and
post-selection picked out a set of ontic states that was
independent of the nature of the intermediate measurement, then
the probability assigned by the ABL rule to a projector could also
be interpreted as the probability assigned to it by these ontic
states.  But since the latter probabilities are required to
satisfy the algebraic conditions in a MNHVT, the violation of
these conditions would be a proof of contextuality. However, the
set of ontic states picked out by the PPS need not be independent
of the nature of the intermediate measurement in general.  To see
this, note that a measurement in an HVT
need not be modelled simply by a Bayesian updating of
one's information, but may also lead to a disturbance of the ontic
state, and the nature of this disturbance may depend on the nature
of the intermediate measurement. Consequently, a PPS paradox is
not itself a proof of contextuality. This is discussed in more
detail in Ref.\ \cite{LeiSpekConf}.

Despite these considerations, the main aim of this letter is to show that there
{\textit is} a connection between PPS paradoxes and
contextuality, but it is significantly more subtle than one might
have thought.

\begin{Theorem}
\label{MainThe} For every logical PPS paradox within the standard
framework for which the pre- and post-selection projectors are
non-orthogonal, there is an associated proof of the impossibility
of an MNHVT that is obtained by considering all the measurements
defined by the PPS paradox -- the pre-selection measurement, the
post-selection measurement and the alternative possible
intermediate measurements -- as alternative possible measurements
at a single time.
\end{Theorem}

Our proof of this theorem generalizes the argument presented for
the 3-box paradox.  We begin with two lemmas and a corollary.

\begin{Lemma} If $\Pi_{\text{pre}}$, $\Pi_{\text{post}}$, $P$ are projectors satisfying
$\Pi_{\text{post}}(I-P)\Pi_{\text{pre}}=0$, then there exists a
pair of orthogonal projectors $Q$ and $R$ such that $I-P=Q+R$
where $\Pi_{\text{pre}}R=0$ and $\Pi_{\text{post}}Q=0$.
\end{Lemma}

\begin{proof}
Let $R\equiv (I-P)\wedge (I-\Pi_{\text{pre}})$, where $P\wedge Q$
denotes the projector onto the intersection of the subspaces
associated with $P$ and $Q$. This clearly satisfies
$\Pi_{\text{pre}}R=0$. Moreover, since $R$ is a subspace of $I-P$,
the projector $Q\equiv (I-P)-R$ is orthogonal to $R$ and satisfies
$I-P=Q+R$.  Finally, $\Pi_{\text{post}}(I-P)\Pi_{\text{pre}}=0$
entails that $\Pi_{\text{post}}$ is orthogonal to the projector
onto ${\text{ran}}((I-P){\text{ran}}(\Pi_{\text{pre}}))$, where
${\text{ran}(X)}$ denotes the range of $X$.  But this projector is
simply $(I-P)-(I-P)\wedge (I-\Pi_{\text{pre}})=Q$. Thus,
$\Pi_{\text{post}}Q=0$ is satisfied.
\end{proof}

\begin{Lemma}
If under a pre-selection of $\Pi_{\text{pre}}$ and a
post-selection of $\Pi_{\text{post}}$, the projector $P$ receives
probability 1 in a measurement of some PVM $\mathcal{E}$, then in
an MNHVT, if $\Pi_{\text{pre}}$ and $\Pi_{\text{post}}$ are
assigned probability 1 by some ontic state $\lambda$, $P$ is also
assigned probability 1 by the ontic state $\lambda$. Succintly, if
$p(P|\Pi _{\text{pre}},\Pi _{\text{post}},\mathcal{E})=1$ and
$p(\Pi_{\text{pre} }|\lambda )=p(\Pi_{\text{post}}|\lambda )=1$,
then $p(P|\lambda )=1$.
\end{Lemma}

\begin{proof}
If $p(P|\Pi _{\text{pre}},\Pi _{\text{post}},\mathcal{E})=1$, then
by the ABL\ rule
\begin{equation*}
\frac{\text{Tr}(\Pi _{\text{pre}}P\Pi
_{\text{post}}P)}{\text{Tr}(\Pi _{ \text{pre}}P\Pi
_{\text{post}}P)+\text{Tr}(\Pi _{\text{pre}}(I-P)\Pi _{\text{
post}}(I-P))}=1
\end{equation*}
which implies that $\text{Tr}(\Pi _{\text{pre}}(I-P)\Pi
_{\text{post}}(I-P))=0$, and since Tr$(A^{\dag }A)=0$ implies that
$A=0,$ it follows that $\Pi _{\text{post}}(I-P)\Pi
_{\text{pre}}=0$. It then follows from lemma 1, that $I-P$ can be
decomposed into a sum of projectors $R$ and $Q$ which are
orthogonal to $\Pi _{\text{pre}}$ and $\Pi _{\text{post}}$
respectively.

Given this orthogonality, for any $\lambda $ in an MNHVT that
yields $p(\Pi _{\text{pre}}|\lambda )=1$, and $p(\Pi
_{\text{post}}|\lambda )$, we have $p(Q|\lambda )=p(R|\lambda
)=0$.  It then follows from the algebraic conditions that
$p(P|\lambda )=1-p(I-P|\lambda ) =1-p(Q+R|\lambda ) =1-p(Q|\lambda
)-p(R|\lambda ) =1$.
\end{proof}

\begin{Corollary}
If $p(P|\Pi _{\text{pre}},\Pi _{\text{post}},\mathcal{E})=0$ and
$p(\Pi_{\text{pre} }|\lambda )=p(\Pi_{\text{post}}|\lambda )=1$,
then $p(P|\lambda )=0$.
\end{Corollary}
\begin{proof} $p(P|\Pi _{\text{pre}},\Pi _{\text{post}},\mathcal{E})=0$ implies
$p(I-P|\Pi _{\text{pre}},\Pi _{\text{post}},\mathcal{E})=1$, which
by lemma 2 implies $p(I-P|\lambda )=1$.  It then follows from the
algebraic constraints that $p(P|\lambda )=0$.
\end{proof}

\begin{proof}[Proof of theorem] \ By the assumption that the PPS
projectors are nonorthogonal, there exist ontic states $\lambda $
such that $p(\Pi _{\text{pre}}|\lambda )=p(\Pi _{\text{
post}}|\lambda )=1.$ \ This, together with lemma 2 and its
corollary, implies that whatever probability assignments to
$\{P\}$ arise from the ABL rule also arise in any MNHVT as the
probability assignment to $\{P\}$ for those ontic states $\lambda
$ yielding $p(\Pi _{\text{pre}}|\lambda )=p(\Pi
_{\text{post}}|\lambda )=1$. \ Since, by the assumption of a
logical PPS paradox, the ABL probabilities violate the algebraic
conditions, it follows that the probabilities conditioned on this
$\lambda$ in an MNHVT also violate the algebraic conditions.
However, probability assignments in an MNHVT must satisfy these
conditions, therefore an MNHVT is ruled out.
\end{proof}

A question that has not been addressed in the present paper is
whether the existence of logical PPS paradoxes in a theory
\emph{implies} measurement contextuality. To answer this question,
one must characterize each of these features in a
theory-independent manner, and examine whether every theory that
exhibits the former also exhibits the latter. \ For an attempt to
provide an operational characterization of contextuality, see
\cite{SpekkensContPMT}. \ No attempt at providing an operational
characterization of logical PPS paradoxes has yet been made, but
in \cite{Kirk3box}, Kirkpatrick has proposed an analogue of the
3-box paradox in the context of a model with playing cards
\cite{Kirkpatrick}, and we have proposed a similar analogue in the
context of a simple partitioned-box model in \cite{LeiSpekConf}.
These models are measurement noncontextual by the definition of
\cite{SpekkensContPMT}.  Thus if one agrees that either the 3-box
paradox of Kirkpatrick or that of \cite{LeiSpekConf} is indeed a
logical PPS\ paradox, then one can conclude that logical PPS
paradoxes \emph{do not }imply contextuality. \

Nonetheless, the toy theory of \cite{SpekkensToy}, which is in
many respects more similar to quantum theory than the models of
\cite{Kirkpatrick} or \cite{LeiSpekConf}, seems
unable to reproduce the logical PPS paradoxes. Moreover, one of
the most conspicuous phenomena that this toy theory fails to
reproduce is contextuality (where again we appeal to the
operational definition of contextuality provided in
\cite{SpekkensContPMT}). All of this suggests that there may be a
natural set of conditions that quantum theory, classical
probability theory with Bayesian updating and the toy theory of \cite{SpekkensToy}
satisfy, but that Kirkpatrick's model and the partitioned box
model of \cite{LeiSpekConf} do \emph{not} satisfy, under which
logical PPS paradoxes can only arise if there is
contextuality. \ Further investigations into these issues are
required.

\begin{acknowledgements}
We would like to thank Ernesto Galv\~{a}o, Terry Rudolph, and Alex
Wilce for helpful comments.
\end{acknowledgements}

\bibliography{contprepost_final}

\end{document}